\documentclass[twocolumn,prb,nofootinbib,showpacs]{revtex4-1}
\usepackage{graphicx}
\usepackage{hyperref}
\usepackage{amsfonts}
\usepackage{amsmath,amssymb}
\usepackage{bm}

\begin{document}

\title{Spin polarization of Majorana zero modes and topological quantum phase transition in semiconductor Majorana nanowires}
\author{Tudor D. Stanescu$^1$}
\author{Sumanta Tewari$^2$}
\affiliation{$^1$Department of Physics and Astronomy, West Virginia University, Morgantown, WV 26506, USA\\
 $^2$Department of Physics and Astronomy, Clemson University, Clemson, SC 29634, USA}

\begin{abstract}
A number of recent works have discussed the issue of spin polarization of a Majorana zero mode in condensed matter systems. Here we show that the spin polarization density of a Majorana zero mode, computed as an average of the spin operator over its wave function, vanishes everywhere. A single non-degenerate Majorana zero mode, therefore, does not couple to an applied magnetic field, except via hybridization with higher energy excited states (if present), which may perturb its wave function. If `spin' is defined by considering only the particle components of the wave function, as has been done in some recent works, Majorana zero modes do have a non-zero spatial profile of this quantity, measurable in scanning tunneling microscopy (STM) experiments. However, if such a quantity is measured in spin-resolved tunneling experiments (without spatial resolution), we show that it cannot be used as a unique signature of Majorana zero modes in the topologically non-trivial phase. As a byproduct, we show that in spatially inhomogeneous systems, accidental zero energy modes, which for all practical purposes behave as Majorana zero modes (including giving rise to a zero bias conductance peak of height $\sim 2e^2/h$), can appear with increasing magnetic field even in the absence of a topological quantum phase transition (TQPT). But only after gap closing and the associated TQPT, the modes are localized near the system edges, resulting in the maximum topological protection. In the light of these considerations, demonstrating the nonlocal character of the topologically-protected Majorana pair and its emergence {\em after} the systems undergo a TQPT, become critical tasks for the ongoing experimental search for Majorana bound states in condensed matter systems.
\end{abstract}

\maketitle

\maketitle

\section{Introduction}

Topological superconductors are defined as systems with a well defined spectral gap to fermionic excitations in the bulk but topologically protected gapless excitations on the surface \cite{Read_Green_2000,Kitaev_2001,Nayak_2008}. Due to the superconducting particle-hole symmetry, the second quantized operators for Bogoliubov excitations in a gapped spinless superconductor, which is a prototype of a topological superconductor \cite{Read_Green_2000,Kitaev_2001}, satisfy the property $\gamma^{\dagger}_{E}=\gamma_{-E}$. It follows that the gapless zero modes on the surface, should they exist,  satisfy the property $\gamma^{\dagger}_0=\gamma_0$, implying particles identifiable with their own anti-particles, first proposed by E. Majorana in 1937 in the context of high energy physics \cite{Majorana}. In the context of condensed matter, the gapless zero energy Bogoliubov excitations, known as Majorana zero modes, emerge as localized zero-energy quasiparticles in topological superconductors bound to defects of the order parameter such as vortices and sample edges. Aside from being fascinating non-elementary particles, in two-dimensional systems Majorana zero modes obey a special type of braiding statistics known as non-Abelian statistics, which is useful in implementing a fault-tolerant topological quantum computer \cite{Kitaev_2001,Nayak_2008}.
These emergent excitations are said to be topologically protected, in the sense that their existence on the surface and other interesting properties are insensitive to perturbations, so long as the system remains gapped in the bulk.

While Majorana zero modes (also sometimes called Majorana bound states (MBSs) or simply Majorana fermions (MFs)) have not yet been conclusively found in experiments, they have been theoretically shown to exist in low dimensional spinless $p$-wave superconducting systems \cite{Read_Green_2000,Kitaev_2001}, as well as other systems which are similar to them \cite{Fu_2008,Zhang_2008,Sato_2009,Sau,Annals,Alicea,Long-PRB,Roman,Oreg,Stanescu}.
In particular, the semiconductor heterostructure scheme, involving a spin-orbit coupled semiconductor in proximity to a s-wave superconductor and an externally applied Zeeman field \cite{Sau,Annals,Alicea,Long-PRB,Roman,Oreg,Stanescu}, has motivated tremendous experimental efforts with a number of recent works claiming to have observed experimental signatures consistent with the existence of Majorana zero modes \cite{Mourik_2012,Deng_2012,Das_2012,Rokhinson_2012,Churchill_2013,Finck_2013}; for a review see Ref.~[\onlinecite{Stanescu_2013}].

Zero energy Majorana bound states in topological superconductors can be viewed as Andreev bound states \cite{Andreev} with an equal admixture of electron- and hole-like components of the same spin. The second quantized Majorana operator, thus, creates and destroys an equal amount of any physical quantity associated with these operators, and the excitation carries zero average charge and spin. Put another way, MBSs, by construction, cannot carry a non-zero value of an internal quantum number, e.g., charge or spin, because if they did, the field that couples to these quantum numbers (e.g., Zeeman field) would be able to remove MBSs locally, which goes against the concept of topological protection \cite{Kitaev_2001}. Despite this, a number of recent works \cite{Simon, Kawakami, Schaffer, Schaffer_2} have proposed the existence of a spin-polarization associated with a single MBS in condensed matter systems.

In this paper we first show that the spin polarization of a MBS, computed as an average of the spin operator over its wave function, vanishes everywhere. An applied Zeeman field therefore has no effect on an isolated MBS, consistent with the concept of topological protection. Not only does the Zeeman field have no effect on the energy of Majorana zero energy state, even the Majorana wave function is unaffected by the Zeeman field provided the gap to the higher energy excited states is large enough (The change in the Majorana wave function by an applied Zeeman field occurs via hybridization with the higher energy excited states, and thus disappears for a truly `stand-alone' MBS). If spin is defined, however, by taking into account only the particle components of the Majorana wave function, as has been done in the recent works \cite{Simon, Kawakami, Schaffer, Schaffer_2}, we show that Majorana zero modes do have a non-zero spatial profile of this quantity, and it can be measured in scanning tunneling microscopy (STM) experiments. However, if this quantity is measured in spin-resolved tunneling spectroscopy (without spatial resolution), we show that it cannot be used as a unique signature of Majorana bound states in the topological superconducting phase of semiconductor-superconductor heterostructures. Our conclusion is that although the `spin polarization' can be probed by spin-resolved tunneling experiments (such as STM and tunneling spectroscopy from the ends) no extra information can be gleaned by spin-resolved tunneling (beyond what is gained by spin-unresolved tunneling) that can help us discriminate between zero energy states in the topologically trivial and non-trivial regimes in the parameter space.

As a byproduct of this work we also show that, in spatially inhomogeneous systems, MBSs can appear with increasing Zeeman field even in the absence of a topological quantum phase transition (TQPT). When the system is still topologically trivial, a regular low energy subgap state near a soft boundary can nucleate two spatially separated zero energy states which (for all practical purposes) behave as Majorana zero modes. So long as the system remains topologically trivial, these Majorana bound states are localized inside the smooth confinement region, while in the topologically non-trivial superconducting phase (i.e., with Zeeman field larger than the critical field required for TQPT) the two  Majorana states are localized near the ends of the wire. In light of this finding we conclude that demonstrating the nonlocal character of the topologically-protected MBS pair and its emergence \textit{after} the system undergoes a TQPT, become critical tasks for the ongoing experimental search for MBSs in solid state structures. In particular, we conjecture that observing a zero-bias conductance peak (of height $\sim 2e^2/h$ ) that sticks to zero energy for a certain range of Zeeman fields \textit{does not} represent a unique signature of the topologically protected Majorana bound states (because such a signature can also appear in the topologically trivial phase, see Fig.~8).

Below in Section II we discuss the issue of spin polarization of MBSs, clearly distinguishing between average spin computed with respect to the full Majorana wave function and another quantity (also sometimes called `spin polarization' of MBSs \cite{Simon, Kawakami, Schaffer, Schaffer_2}) where the average is computed with respect to only the particle components of the wave function. In Section III, we discuss and distinguish MBSs that appear in the topologically non-trivial phase of a semiconductor-superconductor heterostructure with accidental zero energy states in the topologically trivial phase that can also be considered as a pair of MBSs. In Section IV we discuss if `spin polarization' of a MBS (understood as an average with respect to the particle part of the wave function) can help discriminate between the topologically trivial and non-trivial zero energy states in semiconductor-superconductor heterostructures.  We summarize and conclude in Section V.

\section{Spin polarization of Majorana bound states}

Our main goal is to answer the following basic questions: in what sense can one talk about the spin polarization of a Majorana bound state and what are the observable physical implications of the existence of such a property? To answer these questions, we start with the basic equation that describes the quasiparticle dynamics in a superconductor at the mean-field level, the Bogoliubov-de Gennes (BdG) equation, then we discuss the concept of Majorana `spin density' and discuss its possible observable manifestations.

\subsection{Bogoliubov-de Gennes formalism and microscopic model}

The (time independent) BdG  equation describing the mean-field dynamics of quasiparticles in a superconductor  has the generic form
\begin{equation}
{\cal H}_{\rm BdG} ~\!\psi_n = E_n \psi_n,  \label {HBdG}
\end{equation}
where $n=0, \pm1, \pm2, \dots$ is an integer quantum number that labels the quasiparticle energies $E_n$, which, as a consequence of particle-hole symmetry, satisfy the property $E_{-n}=-E_n$ and the eigenvectors $\psi_n$ are 4-component spinors, $\psi_n = (u_{n\uparrow}, u_{n\downarrow}, v_{n\uparrow}, v_{n\downarrow})^T$. The BdG Hamiltonian can be expressed in terms of the first-quantized Hamiltonian ${\cal H}$ of the (normal state) system as ${\cal H}_{\rm BdG} = \frac{1}{2}({\cal H} - {\cal H}^T)\tau_0 +  \frac{1}{2}({\cal H} + {\cal H}^T)\tau_x + \tau_y \sigma_y {\bm \Delta}$, where $\tau_\mu$ and $\sigma_\mu$ are Pauli matrices associated with the particle-hole and spin degrees of freedom, respectively, and ${\bm \Delta}$ is the pairing potential matrix. For concreteness, we will describe a semiconductor wire -- superconductor hybrid system using a simple tight-binding model of $N_y$ parallel coupled chains. The superconductor is not described explicitly (i.e. the corresponding degrees of freedom were already integrated out \cite{Stanescu_2013}, but it induces a local (s-wave) pairing potential  $\Delta$ in the chains, so that ${\bm \Delta}_{ij} = \Delta \delta_{ij}$, where $i=(i_x, i_y)$ and $j=(j_x, j_y)$ are site labels satisfying the conditions $1\leq i_x, j_x \leq N_x$ and  $1\leq i_y, j_y \leq N_y$. The (second quantized) Hamiltonian $H=\sum_{i.j}\sum_{\sigma,\sigma^\prime} c_{i\sigma}^\dagger {\cal H}_{i\sigma~\!\! j \sigma^\prime} ~\!c_{j\sigma^\prime}$ describing the normal state of the coupled chains has the general form
\begin{equation}
H=H_0 +H_{\rm SOI} + H_Z+ V,  \label{Hchains}
\end{equation}
where the the first term corresponds to nearest neighbor hopping along and between the chains, the second describes spin-orbit coupling, the third describes the Zeeman coupling to an external magnetic field, and the fourth includes confining and disorder potential terms. Explicitly,  we have
\begin{eqnarray}
&\!&H_0 = -t_x\sum_{i, \delta_x, \sigma}\! c_{i+\delta_x \sigma}^\dagger c_{i \sigma} - t_y\sum_{i, \delta_y, \sigma}\! c_{i+\delta_y \sigma}^\dagger c_{i \sigma} -\mu\sum_{i, \sigma}\! c_{i\sigma}^\dagger c_{i\sigma}, \nonumber \\
&\!&H_{\rm SOI} = \frac{i}{2} \sum_{i, {\bm \delta}}\left[\alpha c_{i+\delta_x}^\dagger \sigma_y c_i - \alpha_y c_{i+\delta_y}^\dagger \sigma_x c_i +{\rm h.c.}\right], \nonumber \\
&\!&H_Z = \Gamma\sum_i c_i^\dagger \sigma_x c_i, \label{Hz}\\
&\!&V=\sum_{i, \sigma} V_c(i) c_{i\sigma}^\dagger c_{i\sigma} + \sum_{i, \sigma, \sigma^\prime} W_{\sigma \sigma^\prime}(i) c_{i\sigma}^\dagger c_{i\sigma^\prime}, \nonumber
\end{eqnarray}
where $t_x$ and $t_y$ are intra- and inter-chain nearest neighbor hopping matrix elements, respectively, $\mu$ is the chemical potential, $\alpha$ and $\alpha_y$ are the longitudinal and transverse Rashba coefficients, respectively,  $\Gamma$ is the Zeeman splitting, $V_c$ is a  confining potential, $W$ is a spin-dependend disorder potential and we have used the spinor notation $c_i = (c_{i\uparrow}, c_{i\downarrow})^T$.

As a consequence of the intrinsic particle-hole redundancy of the BdG theory, the Hamiltonian that describes the dynamics of the quasiparticles has particle-hole symmetry, i.e. it satisfies the following relation
\begin{equation}
{\cal H}_{\rm BdG} = -\tau_x {\cal H}_{\rm BdG}^T\tau_x.        \label{PHsymm}
\end{equation}
This implies that the eigenvectors from Eq. (\ref{HBdG}) corresponding to $n$ and $-n$ are not independent. Specifically, we have $\psi_{-n} = e^{i\varphi}\tau_x\psi_n^*$, where $e^{i\varphi}$ is a constant phase factor. In terms of particle and hole components, this means that the particle component of the eigenvector corresponding to energy $E_n$ is related to the hole component of the eigenvector corresponding to $-E_n$,
\begin{equation}
v_{-n\sigma} = u^*_{n\sigma} e^{i\varphi}.  \label{vnun}
\end{equation}
For a Majorana bound state, which is a solution of  Eq. (\ref{HBdG}) corresponding to $n=0$, particle-hole imposes the constraint $v_{0\sigma} = u^*_{0\sigma} e^{i\varphi}$. Consequently, the spinor describing a zero-energy Majorana state has the following generic form
\begin{equation}
\psi_0 =(u_{0\uparrow}, u_{0\downarrow}, u_{0\uparrow}^*e^{i\varphi}, u_{0\downarrow}^*e^{i\varphi})^T.  \label{psi0}
\end{equation}
In this equation, $u_{0\sigma} \equiv u_{0\sigma}(i_x, i_y)$ are functions of position and $e^{i\varphi}$ is a constant phase.
We note that Eq. (\ref{psi0}) requires only particle-hole symmetry and the vanishing of the energy, $E_0=0$, i.e. it  holds in the presence of any type of perturbation (e.g., spin-dependent disorder) as long as the two requirements are satisfied. We also note that in the presence of additional symmetries, e.g., chiral symmetry \cite{Schnyder,Tewari_Chiral}, which is realized in the absence of transverse Rashba coupling \cite{Tewari_Stanescu_Chiral}, $\alpha_y=0$, Eq. (\ref{psi0}) can be further simplified \cite{Dumitrescu_Hidden}.

\subsection{Spin density operator and Majorana spin density}

To clarify the concept of `Majorana spin density', let us start with the expression of the spin density operator in the BdG formalism. Let us consider the (first quantized) spin operator ${\bm S}= \hat{e}_x S_x + \hat{e}_y S_y +\hat{e}_z S_z$, where $\hat{e}_\mu$ are unit vectors in a Cartesian coordinate system and $S_\mu =\frac{\hbar}{2}\sigma_\mu$ are operators for the corresponding spin components. The (second quantized) spin density operator can be expressed as ${\bm S}_i = c_{i\sigma}^\dagger {\bm S}_{\sigma\sigma^\prime}c_{i\sigma^\prime} = \frac12[c_{i\sigma}^\dagger {\bm S}_{\sigma\sigma^\prime}c_{i\sigma^\prime} - c_{i\sigma} ({\bm S}^T)_{\sigma\sigma^\prime}c_{i\sigma^\prime}^\dagger]$, where, for simplicity, we have used the same symbol for the first and second quantized spin operators. Introducing the four-component  spinor notation $\hat{\psi}_i = (c_{i\uparrow},  c_{i\uparrow}, c_{i\uparrow}^\dagger, c_{i\downarrow}^\dagger)^T$, we can write the spin density operator as  ${\bm S}_i = \frac12 \hat{\psi}_i^\dagger {\cal S} \hat{\psi}_i$, where the ${\cal S}$ is the spin operator in the BdG format,
\begin{equation}
{\cal S} = \left(\begin{array}{cc}
{\bm S} & 0 \\
0 & -{\bm S}^T
\end{array}\right).  \label{SBdG}
\end{equation}
Given the wave function $\psi_n(i)$ of a generic Bogoliubov quasiparticle, we can write the corresponding spin density as ${\bm s}_n(i) =\frac12   \psi_n^\dagger(i) {\cal S} \psi_n(i)$. If we consider now the case of a zero-energy Majorana state, Eqns. (\ref{psi0}) and (\ref{SBdG}) imply
\begin{equation}
{\bm s}_0(i)=\frac12 \psi_0^\dagger(i) {\cal S} \psi_0(i)=0. \label{s0}
\end{equation}
We note that a similar conclusion can be reached concerning the charge density of the Majorana state.
In other words, {\em the charge and spin densities of a zero-energy Majorana state are identically zero}. We emphasize that these are local (rather than global) properties: not only the total spin and charge of the Majorana vanish, but the corresponding densities are identically zero.

Our conclusion so far, namely that zero-energy Majorana states do not carry spin and charge, is certainly not surprising. However, the key questions concern the observable physical consequences of this property. More specifically, what we want to understand is  i) whether or not a Majorana bound state couples to an external magnetic field and ii) if there is any unique signature of Majorana bound states in spin-resolved tunneling experiments. Furthermore, our analysis did not distinguish between Majorana bound states that emerge in a topological superconducting state (e.g., as zero-energy states localized near the ends of a superconducting wire) and regular zero-energy Bogoliubov quasiparticles, which may occur in a topologically trivial superconductor. Is there any spin-related property that can be used to discriminate between these types of zero-energy states?  In the subsequent sections of this article we will discuss these issues based on a numerical analysis of the effective model for a system of coupled superconducting chains given by Eqns. (\ref{Hchains}) and (\ref{Hz}).

Returning to the spin density, let us remark that the vanishing of ${\bm s}_0(i)$ is due to an exact cancellation between the particle and hole contributions to this quantity. Taken separately, these contributions are, in general, nonzero. Hence, the natural question is, can we ascribe any physical significance to the particle (or hole) contribution to the spin density? More specifically, let us define the quantity
\begin{equation}
\langle{\bm S}\rangle(i) = \frac12 u_{0\sigma}^*(i){\bm S}_{\sigma\sigma^\prime}u_{0\sigma^\prime}(i).  \label{Si}
\end{equation}
Is there a measurement that probes this quantity? The answer turns out to be affirmative, as has been pointed out in the recent works \cite{Simon, Kawakami, Schaffer, Schaffer_2}, as
spin-resolved local density of states (LDOS) and spin-resolved tunneling spectroscopy are related to this quantity.

In the remainder of this article we will call the quantity $\langle{\bm S}\rangle(i)$ given by Eq. (\ref{Si}) `{\em spin density}'. We emphasize, however, that this is an abuse of language, as $\langle{\bm S}\rangle(i)$ represents only the particle component of the spin density ${\bm s}_0(i)$, which is identically zero. Below, we identify different contexts in which one of these concepts of spin density or the other are relevant. We focus on the question concerning whether or not these concepts are helpful when trying to distinguish between topological Majorana bound states  and topologically trivial zero-energy Andreev bound states.

\section{Majorana bound states versus topologically trivial zero-energy states}

Discriminating between the Majorana bound states predicted to emerge in a topologically nontrivial superconductor and certain trivial  zero-energy Andreev bound states represents a crucial component of the ongoing search for Majorana quasiparticles. Below we address the question if the concept of spin density shed any light on this problem.

\begin{figure}[tbp]
\begin{center}
\includegraphics[width=0.48\textwidth]{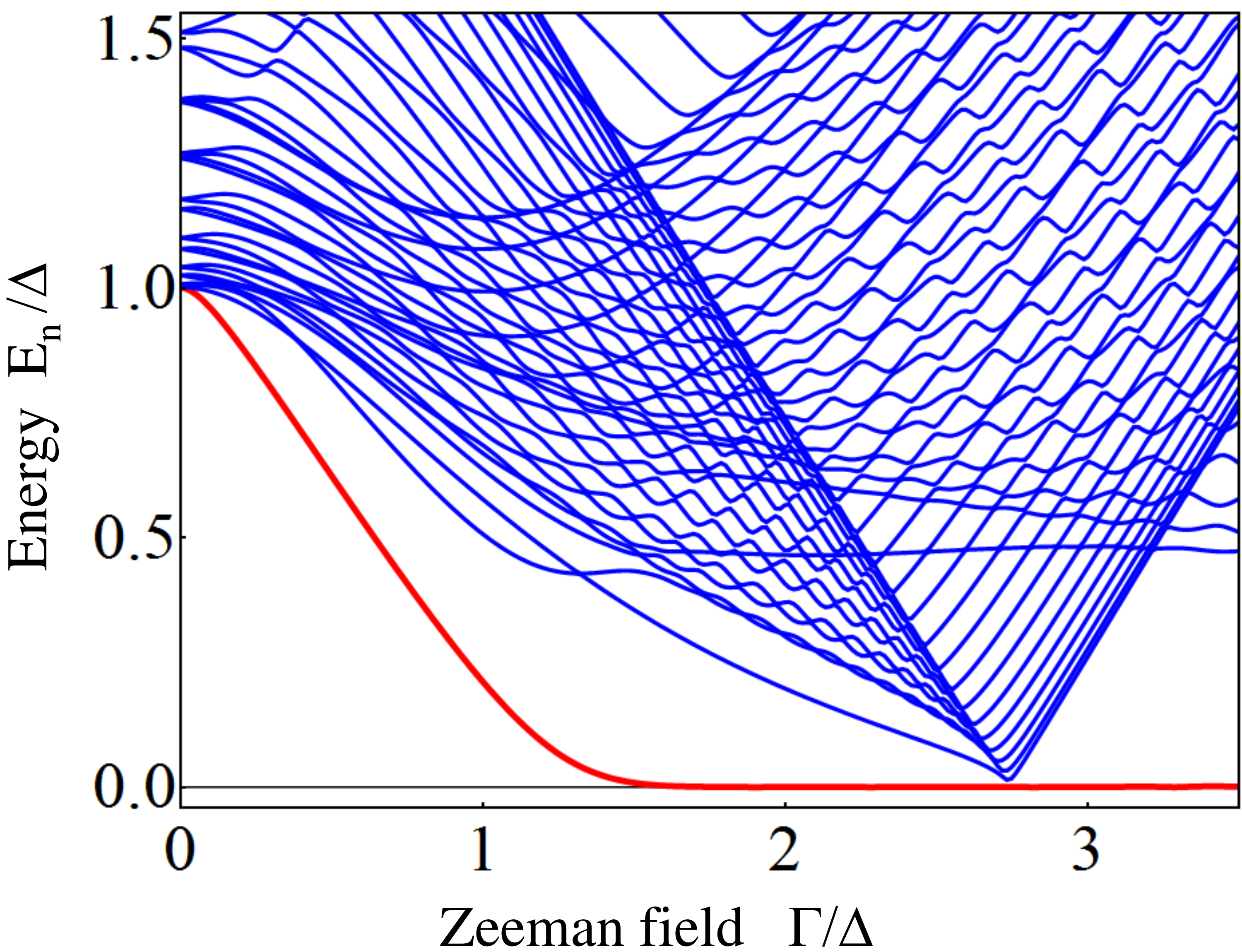}
\end{center}
\caption{Dependence of the low-energy BdG spectrum on the applied Zeeman field for a system of superconducting chains with smooth confinement. Only the positive energies are shown. The confinement potential profile is shown in Fig. \ref{Fig2}. Note the vanishing of the bulk gap (blue lines) at the critical field $\Gamma_c =\sqrt{\mu^2+\Delta^2}\approx 2.7\Delta$, which  signals a topological quantum phase transition. The zero-energy mode extends into the topologically trivial phase $\Gamma<\Gamma_c$.}
\vspace{-4mm}
\label{Fig1}
\end{figure}

\subsection{Trivial zero-energy states in a superconducting wire with smooth confinement}

There are several different scenarios that predict the occurrence of zero-energy states in a topological trivial superconductor, e.g., in the presence of disorder \cite{Liu,Roy} or in a system with smooth confinement \cite{Kells}. To better understand the similarities and the differences between these trivial states and the topological Majorana modes, we consider the chain model given by Eqns. {\ref{Hchains}) and (\ref{Hz}) for a system with hard-wall confinement at one end and smooth confinement at the other end. Specifically, we have
\begin{equation}
V_c(i) =\left\{\begin{array}{l}
V_c^{\rm max}~\frac{\delta+1-i_x}{\delta+1}~~~~~~~~~~{\rm if}~~ 1\leq i_x\leq \delta, \\
0~~~~~~~~~~~~~~~~~~~~~~~~~~\!{\rm if}~~ \delta<i_x\leq N_x, \\
\infty~~~~~~~~~~~~~~~~~~~~~~~~{\rm otherwise}.
\end{array}\right.
\end{equation}
The position dependence of the confining potential is represented schematically in the upper panel of Fig. \ref{Fig2}. The parameters used in the numerical calculation are $\delta=100$, $N_x=400$, and $V_c^{\rm max} = 5\Delta$, where $\Delta$ is the induced superconducting pairing and will represent our unit for energy. The values of the other model parameters used in the calculation are $\mu=2.5\Delta$,  $t_x=50 \Delta$, $t_y=10\Delta$, $\alpha=5\Delta$, and $\alpha_y=\Delta$. Note that, with a nearest neighbor distance along the chains $a_x=10~\!$nm, a separation between neighboring chains $a_y\approx 22~\!$nm and an induced pairing potential $\Delta = 0.4~\!$meV, the hopping parameters correspond to an effective mass $m_{eff}\approx 0.02m_0$, where $m_0$ is the free electron mass, and the Rashba spin-orbit coupling coefficient is $\alpha_R\approx 200~\!$meV\AA, which are typical values for the semiconductors used in experiments.

\begin{figure}[tbp]
\begin{center}
\includegraphics[width=0.48\textwidth]{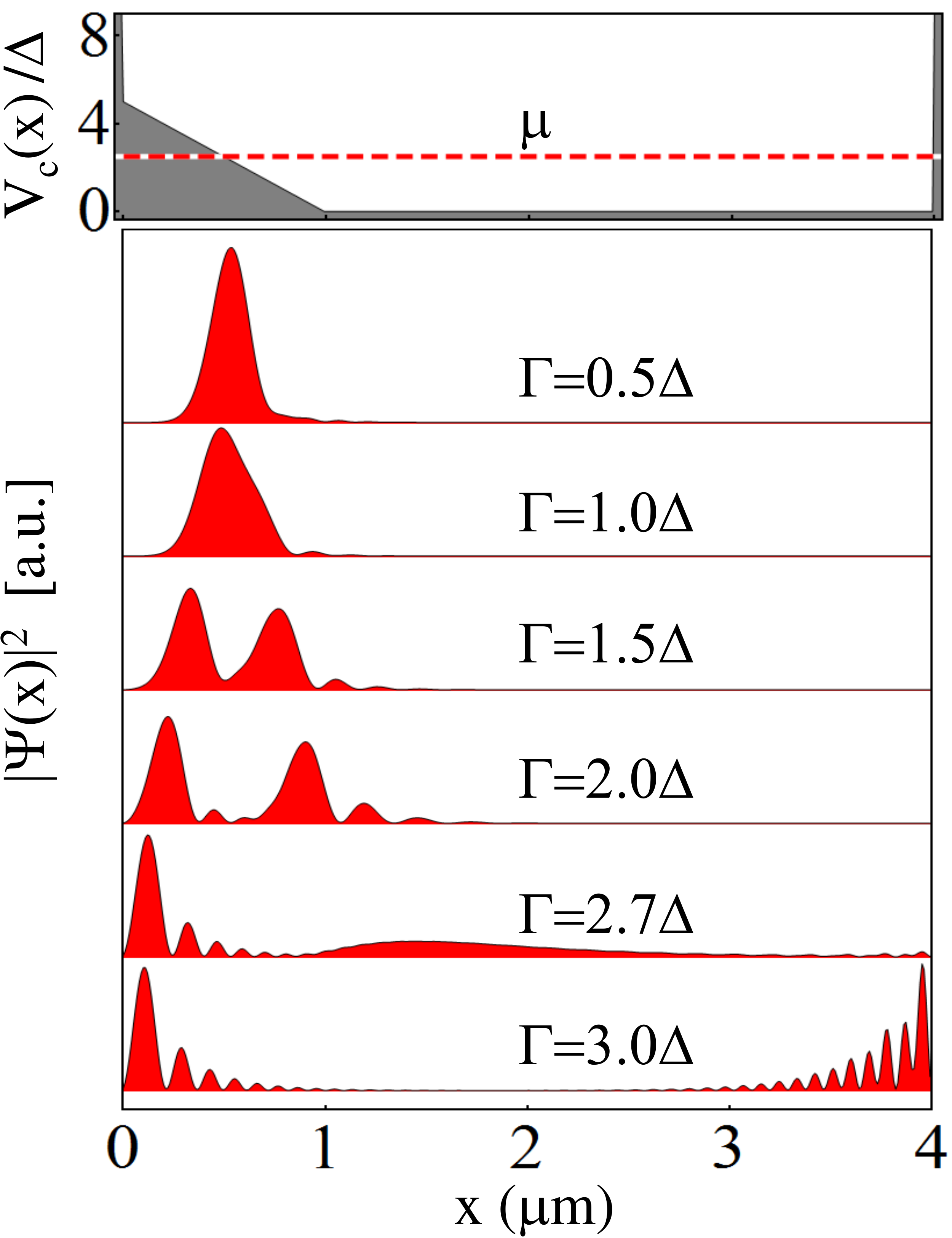}
\end{center}
\caption{Upper panel: Schematic representation of the confining potential profile. The system has smooth confinement at the left end of the chains and hard-wall confinement at the right end. Lower six panels: Evolution of the wave function corresponding to the lowest energy mode in Fig. \ref{Fig1} with the applied Zeeman field. The energy of the mode vanishes for $\Gamma \gtrsim 1.6\Delta$.}
\vspace{-4mm}
\label{Fig2}
\end{figure}

In the absence of an applied magnetic field, the system is in a topologically trivial superconducting state and the corresponding BdG spectrum is characterized by a gap $2\Delta$. To observe a Majorana zero mode, we need to turn on the Zeeman field. More specifically, we expect the Majorana bound states to emerge in the topological superconducting state that obtains for values of the applied field above the critical value $\Gamma_c =\sqrt{\mu^2+\Delta^2}\approx 2.7\Delta$.  The actual dependence of the quasiparticle energies on the applied Zeeman field is shown in Fig. \ref{Fig1}.  The key features are i) the emergence of a zero-energy mode (red line) for $\Gamma \gtrsim 1.6\Delta$ and ii) the vanishing of the bulk gap (blue lines) at $\Gamma=\Gamma_c$. For values of the Zeeman field in the range $1.6\Delta\lesssim \Gamma\lesssim 2.7\Delta$ the system is in a topologically trivial phase but supports a zero-energy mode.

\begin{figure}[tbp]
\begin{center}
\includegraphics[width=0.48\textwidth]{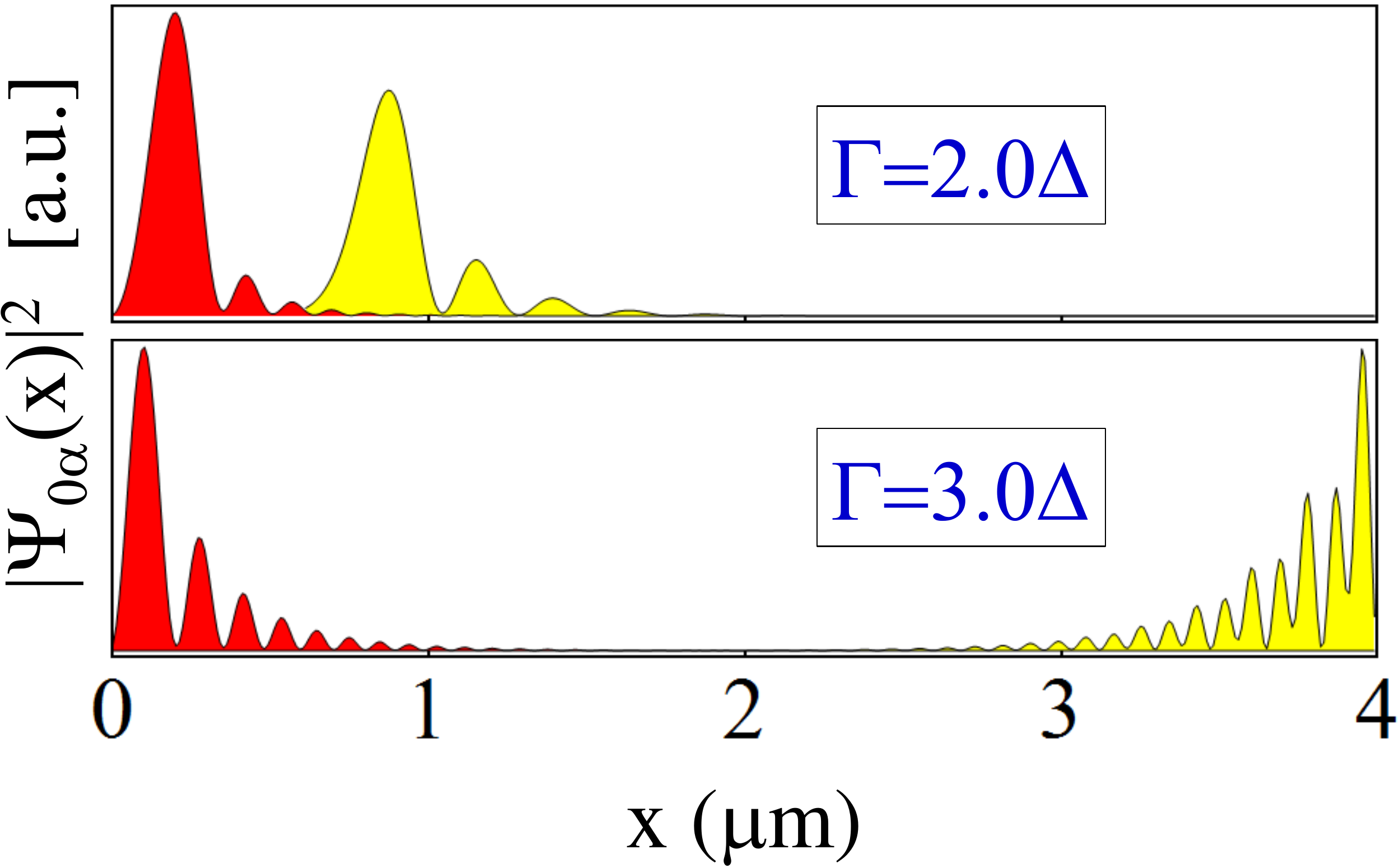}
\end{center}
\caption{Pair of Majorana bound states associated with the zero-energy mode. The corresponding wave functions are $\psi_{01}$ (red filling) and $\psi_{02}$ (yellow filling). In the topologically trivial phase (upper panel) the Majorana bound states are localized inside the smooth confinement region, while in topological superconducting phase (lower panel) the two  Majorana states are localized near the ends of the wire.}
\vspace{-4mm}
\label{Fig3}
\end{figure}

To better understand the nature of the lowest energy mode (red line in Fig. \ref{Fig1}), we calculate the corresponding wave function for different values of the applied field. The results are shown in Fig. \ref{Fig2}. For $\Gamma<\Gamma_c$, i.e. in the topologically trivial phase, the lowest energy mode corresponds to a bound state localized near the softly confined end of the wire. The corresponding wave function is characterized by a single peak at low fields and splits into two separated components as the strength of the Zeeman field increases. As these these components become clearly separated (i.e for $\Gamma \gtrsim 1.6\Delta$), the energy of the mode vanishes. At the critical field $\Gamma_c\approx 2.7\Delta$ one of the components becomes delocalized, then (i.e. inside the topological superconducting phase) the wave function has two components exponentially localized near the two ends of the wire (see Fig. \ref{Fig2}, lowest panel).

To gain further insight, we note that, due to the intrinsic redundancy of the BdG description, the zero-energy mode is in fact double degenerate. The corresponding quasiparticles, which represent a single Dirac fermion, can be viewed as a pair of Majorana modes. The wave functions of the two Majoranas $\psi_{0\alpha}$, with $\alpha=1,2$, have the form given by Eq. (\ref{psi0}) and can be obtained by taking linear combinations of the zero-energy solutions obtained numerically. In the topological regime ($\Gamma>\Gamma_c$), the zero-energy quasiparticles are two Majorana bound states localized near the ends of the wire, as shown in the lower panel of Fig. \ref{Fig3}. By contrast, the zero-energy bound state that emerges in the topologically trivial regime ($\Gamma<\Gamma_c$) consists of two (weakly overlapping) Majorana bound states localized in the region with nonuniform, smoothly varying confinement potential (see Fig. \ref{Fig3}, upper panel).

We emphasize that in a superconductor {\em any zero-energy state} can viewed as a linear combination of two Majorana quasiparticles (or, more generally, an even number of them). However, the stability of the state, i.e. whether or not it becomes gapped when a perturbation is applied to the system, depends on the overlap between these Majoranas. In principle, perfect stability requires the absence of any overlap, which in the case of Majorana bound states emerging in a quasi one-dimensional superconductor corresponds to the ideal limit of infinitely long wires. The realistic, finite length system is discussed below.

\subsection{Ideal versus effective Majorana bound states}

The Majorana bound states localized near the ends of a topological superconducting wire are characterized by wave functions that have exponentially decaying envelopes (see, for example, the lower panel of Fig. \ref{Fig3}). In a finite wire, the exponentially small tails generate a finite splitting of zero mode that oscillates as a function of the applied Zeeman field \cite{Dassarma}. Hence, the lowest energy mode (i.e. the red line in Fig. \ref{Fig1}) has exactly zero energy only at a discrete set of $\Gamma$ values and finite (although very small) energy everywhere else. This is illustrated in the upper panel of Fig. \ref{Fig4}. Note that, qualitatively, there is no difference between the topologically trivial and nontrivial regimes.

The first obvious question is the following: can we actually talk about Majorana bound states for values of the Zeeman field that are different from the nodes of the lowest energy mode? After all, the corresponding excitations are just regular, finite energy Bogoliubov quasiparticles? To answer this question we need a conceptual clarification. Specifically, we introduce a distinction between i) {\em ideal} Majorana bound states and ii) {\em effective} Majorana bound states.
An ideal Majorana bound state (MBS) is a zero-energy state localized near the end of an infinitely long quasi-1D topological superconductor. The ideal MBS does not overlap with its partner at the opposite and, consequently, it is an exact zero-energy eigenstate of the BdG Hamiltonian regardless of how one perturbs the system (as long as the superconducting gap remains open). An effective MBS, on the other hand, has a small (but finite) overlap with its partner and is not an exact eigenstate of the BdG Hamiltonian. However, the states associated with the nearly-zero, lowest-energy mode of the system are linear combinations of these effective MBSs.

\begin{figure}[tbp]
\begin{center}
\includegraphics[width=0.48\textwidth]{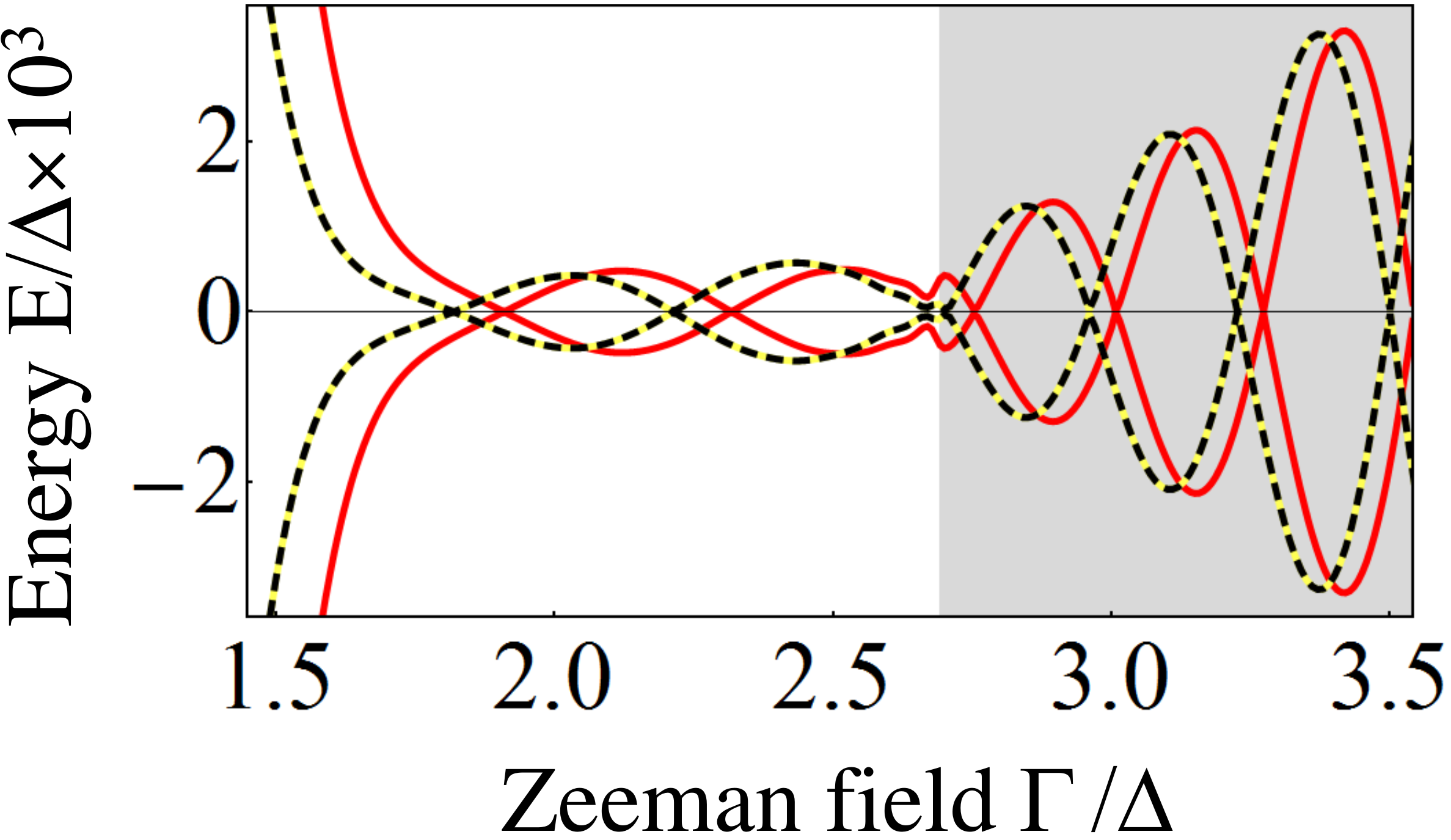}
\end{center}
\caption{Lowest energy mode as function of the applied Zeeman field. The continuous red line is the same as in Fig. \ref{Fig1}, while the dashed black line is obtained by applying an additional local field with $\delta\Gamma_0=0.25\Delta$ (see main text). The gray region corresponds to the topological regime.}
\vspace{-4mm}
\label{Fig4}
\end{figure}

Specifically,  let us consider a state $\psi_+$ of energy $\epsilon>0$ corresponding to the nearly-zero mode (e.g., the red line in Fig. \ref{Fig4}) and its negative energy partner $\psi_-$ of energy $-\epsilon$. The corresponding effective Majorana bound states $\psi_{01}$ and $\psi_{02}$ have the generic form given by Eq. (\ref{psi0}), more specifically, $\psi_{01} = (u_{01}, u_{01}^*e^{i\varphi})^T$ and  $\psi_{02} = (u_{02}, -u_{02}^*e^{i\varphi})^T$, where we used the notation $u_{0i} = (u_{0i\uparrow}, u_{0i\downarrow})^T$. The wave functions of the finite-energy Bogoliubov quasiparticles can be expressed in terms of the effective MBS wave functions as
\begin{eqnarray}
\psi_+ &=& \frac{1}{\sqrt{2}}(\psi_{01} + \psi_{02}), \nonumber \\
\psi_- &=& \frac{1}{\sqrt{2}}(\psi_{01} - \psi_{02}),   \label{psipm}
\end{eqnarray}
In the calculations, we determine $\psi_{\pm}$ by diagonalizing the BdG Hamiltonian, then we obtain the effective MBSs by inverting Eq. (\ref{psipm}), after properly fixing the relative phase between $\psi_{+}$ and $\psi_{-}$. Note that we have $\langle\psi_{0i}|{\cal H}_{\rm BdG}|\psi_{0i}\rangle = 0$, but the matrix element of the Hamiltonian between the two different MBSs is finite as a result of them having a nonzero overlap,  $\langle\psi_{01}|{\cal H}_{\rm BdG}|\psi_{02}\rangle = \epsilon$. We emphasize that this construction can be done for arbitrary finite energy eigenstates of the BdG Hamiltonian, but the resulting Majorana states $\psi_{0i}$ will generally be extended, rather than localized. In other words, the concepts of {\em effective} MBS is meaningful only at low energy, i.e. for $\epsilon$ less that than a certain energy scale associated, for example, with the experimental energy resolution or the inverse timescale for MBS manipulation, a key quantity when considering the braiding of Majorana bound states.

\begin{figure}[tbp]
\begin{center}
\includegraphics[width=0.48\textwidth]{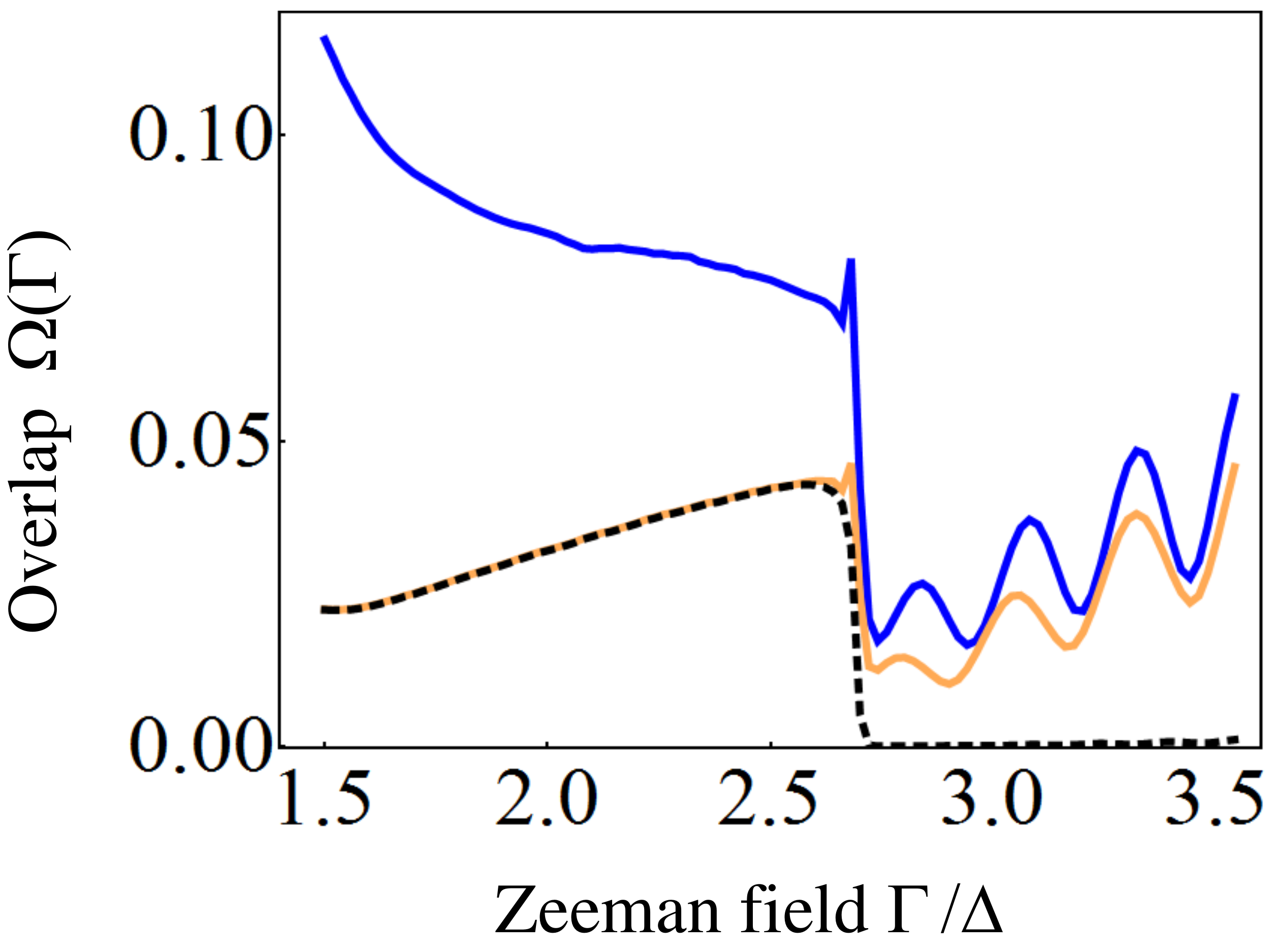}
\end{center}
\caption{Overlap as a function of the applied Zeeman field for a wire with soft confinement. The confinement potential has the same profile as in Fig. \ref{Fig2} but the confinement region is 1.5 $\mu$m long. The total length of the wire and the slope of the confining potential are: $L_x=3.5~\!\mu$m, $\theta = 5\Delta/\mu$m (blue line);  $L_x=3.5~\!\mu$m, $\theta = 2.5\Delta/\mu$m (orange line); $L_x=6.5~\!\mu$m, $\theta = 2.5\Delta/\mu$m (dashed black line).}
\vspace{-4mm}
\label{Fig4bis}
\end{figure}

Above, we have mentioned several times the term `overlap' in relation to a pair of effective MBSs, but without being too specific. Obviously, this cannot refer to the matrix element $\langle\psi_{01}|\psi_{02}\rangle$, which is always zero.  What we actually mean by two MBSs having a nonzero overlap is that there is a certain region where the corresponding wave functions are both nonzero. We can quantify this by defining the following quantity, which we will call the overlap of $\psi_{01}$ and $\psi_{02}$,
\begin{equation}
\Omega = \sum_{i, \sigma} (|u_{01\sigma}(i)||u_{02\sigma}(i)| + |v_{01\sigma}(i)||v_{02\sigma}(i)|). \label{Ovl}
\end{equation}
For a pair of ideal Majorana bound states we have $\Omega=0$. In a finite system, or in the case of topologically trivial nearly-zero energy states the overlap is finite, but typically $\Omega \ll 1$.  Fig. \ref{Fig4bis} shows the dependence of the overlap on the applied Zeeman field for a wire with soft confinement. We note that, as expected, in the topological regime ($\Gamma>\Gamma_c$) the overlap can be made arbitrarily small be increasing the length of the wire.  However, one can also reduce the overlap in the topologically trivial phase by reducing the slope of the confining potential.

As evident from Fig. \ref{Fig4bis}, the size of overlap is, in general,  not a good criterion for distinguishing between the topological and the trivial regimes in finite systems (e.g., the orange curve in Fig. \ref{Fig4bis}). Nonetheless, there is a clear feature associated with the topological quantum phase transition at $\Gamma=\Gamma_c$, namely the sharp drop of $\Omega(\Gamma)$ at the critical field. This feature occurs because the overlap $\Omega$ contains information about the spatial profile of the effective MBSs.  Unfortunately, this information cannot be captured by probes such as  charge (or spin) tunneling into the end of the wire, which show no signature associated with the topological quantum phase transition, as we will show in Section  \ref{SR_tunnelingB}.

We close this section with a comment concerning  the topological quantum phase transition (TQPT) itself. The transition is associated with the vanishing of the bulk gap, as mentioned before. However, in a finite system the bulk gap never really closes.  Again, we have to distinguish between an {\em ideal} TQPT, which corresponds to a point in parameter space (e.g., Zeeman field strength) where the bulk gap is exactly zero and can only take place in an infinite system, and {\em effective} TQPTs, which occur in finite topological superconductors and are associated with a minimum (rather than a zero) of the bulk gap.  Strictly speaking, an effective TQPT is always a crossover, rather than a true phase transition. Nonetheless, in certain situations one can identify a natural energy scale that allows us to consider the bulk gap as ``effectively'' zero  when it becomes smaller that this characteristic energy. In these cases we talk about an effective TQPT, rather than a crossover. Note, however, that having effective MBSs does not necessarily imply that the system has crossed an effective TQPT. These considerations are particularly relevant when studying non-homogeneous superconductors, such as the superconducting wires with soft confinement discussed here or various proposed realizations of Majorana states in trapped ultracold atomic gases.

\begin{figure}[tbp]
\begin{center}
\includegraphics[width=0.48\textwidth]{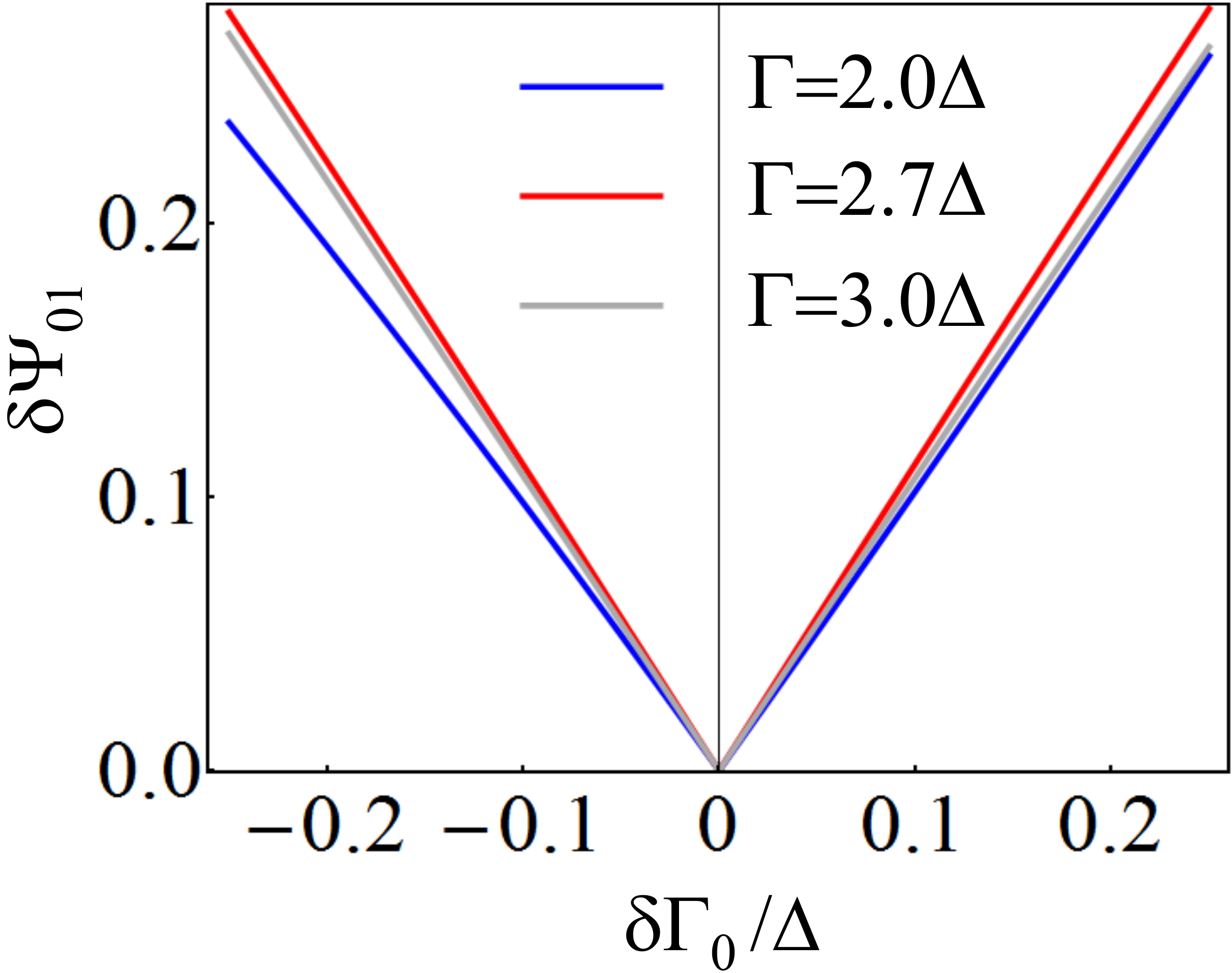}
\end{center}
\caption{Dependence of the change in the Majorana wave function $\psi_{01}$ (see Fig. \ref{Fig3}) on the additional local magnetic field. The change is due to the lowest energy bound state hybridizing with finite energy states that have significant amplitude near the left end of the wire. For small values of the perturbation the slope is inversely proportional to the characteristic energy of these states. Note the similar  slopes associated with different values of $\Gamma$, which indicates that $\psi_{01}$ hybridizes with finite energy states that depend weakly on the Zeeman field (see the nearly horizontal modes in Fig. \ref{Fig1}).}
\vspace{-4mm}
\label{Fig4bisbis}
\end{figure}

\subsection{Coupling of Majorana bound states to external magnetic fields}

Having discussed how the Majorana bound states can emerge in a superconductor wire, we return to our considerations of the concept of spin density associated with these state. We have defined the spin density of a MBS in two different  ways: in terms of the full wave function [see Eq. (\ref{s0})], which gives us a spin density ${\bm s}(i)$ that is identically zero, and in terms of the particle component of the wave function [see Eq. (\ref{Si})], which gives us the nonzero quantity $\langle{\bm S}\rangle(i)$. To clarify the meaning of these quantities, we couple the system to a local magnetic field in the $x$ direction, which is applied to the left end of the wire and generates an additional  Zeeman splitting of the form
\begin{equation}
\delta\Gamma(x) = \delta\Gamma_0~\! e^{-x/d}. \label{dgamma}
\end{equation}
We choose $d=0.4~\!\mu$m, so that the additional Zeeman field is significant only in the region where the leftmost  MBS $\psi_{01}$ is located. The energy of any quasiparticle that possesses a `true' spin polarization is expected to be shifted by an energy proportional to $\delta\Gamma_0$, at least in the low perturbation limit. This is not the case for a Majorana bound state. For an ideal MBS, the energy remains unchanged, $E_0=0$, showing that the bound state carries no spin, hence it does not couple to an external magnetic field. For an effective MBS, there is a small change in the energy due to induced coupling to higher energy states, but no direct coupling to the Zeeman field. As a result, the energy shift depends non-monotonically on $\delta\Gamma_0$ and changes sign as function of the overall Zeeman field $\Gamma$, as illustrated in Fig. \ref{Fig4}. Also note that there is no qualitative difference between the effective MBS emerging in the topological regime and the effective MBS generated in the trivial regime by the soft confinement. Also note that the amplitude of the splitting oscillations is not significantly affected by the local perturbation. These observations hold for local Zeeman fields with arbitrary spatial profiles and arbitrary orientations.  We conclude that a Majorana bound state has no spin polarization and that its spin density is properly described by the quantity ${\bm s}(x) =0$.

In this context, we note that although a local Zeeman field does not affect significantly (or at all, in the ideal case) the energy of  a MBS, it can modify its wave function by coupling it to higher energy states. If $\psi_{01}$ is the unperturbed wave function of the MBS and $\psi_{01}^\prime$ its wave function in the presence of the local perturbation (\ref{dgamma}), we define the change in the Majorana wave function as
\begin{equation}
\delta\psi_{01} =\sqrt{\langle\psi_{01}-\psi_{01}^\prime|\psi_{01}-\psi_{01}^\prime\rangle}.
\end{equation}
Based on second-order perturbation theory one would expect this quantity to depend linearly on the strength $\delta\Gamma_0$ of the perturbation (in the small perturbation limit), with a slope that is inversely proportional to the energy of the finite energy state that couples to the MBS. This is confirmed by the numerical results shown in Fig. \ref{Fig4bisbis}.  Note again that there is no qualitative difference between the topological and the trivial regimes. Also note that the slopes of the curves shown in Fig. \ref{Fig4bisbis} are very similar and are not correlated with the (inverse) bulk gap shown in Fig. \ref{Fig1}. This is due to the fact that the MBS does not couple effectively to most of the bulk states, which have negligible amplitudes in the vicinity of the left end of the wire. Instead, $\psi_{01}$ coupled to certain modes that have a weak dependence on the Zeeman field (in the relevant range) and can be identified as  nearly horizontal, finite energy lines in Fig. \ref{Fig1}.

\section{Signatures of  Majorana bound states in spin-resolved tunneling}

Based on the (lack of) coupling to an external magnetic field, we concluded that Majorana bound states have no spin polarization. Similar considerations involving a local electrostatic potential support the fact that Majorana states carry no charge, hence are characterized by an identically zero charge density. However, a Majorana bound state can be probed by tunneling charge into the end of the wire, the characteristic signature being a zero-bias conductance peak. Similarly, one would expect a nontrivial signature when tunneling spin-polarized electrons into a superconducting wire that host MBSs. Below, we briefly discuss these signatures and whether or not they can help to distinguish between the topological and the trivial superconducting regimes.

\begin{figure}[tbp]
\begin{center}
\includegraphics[width=0.48\textwidth]{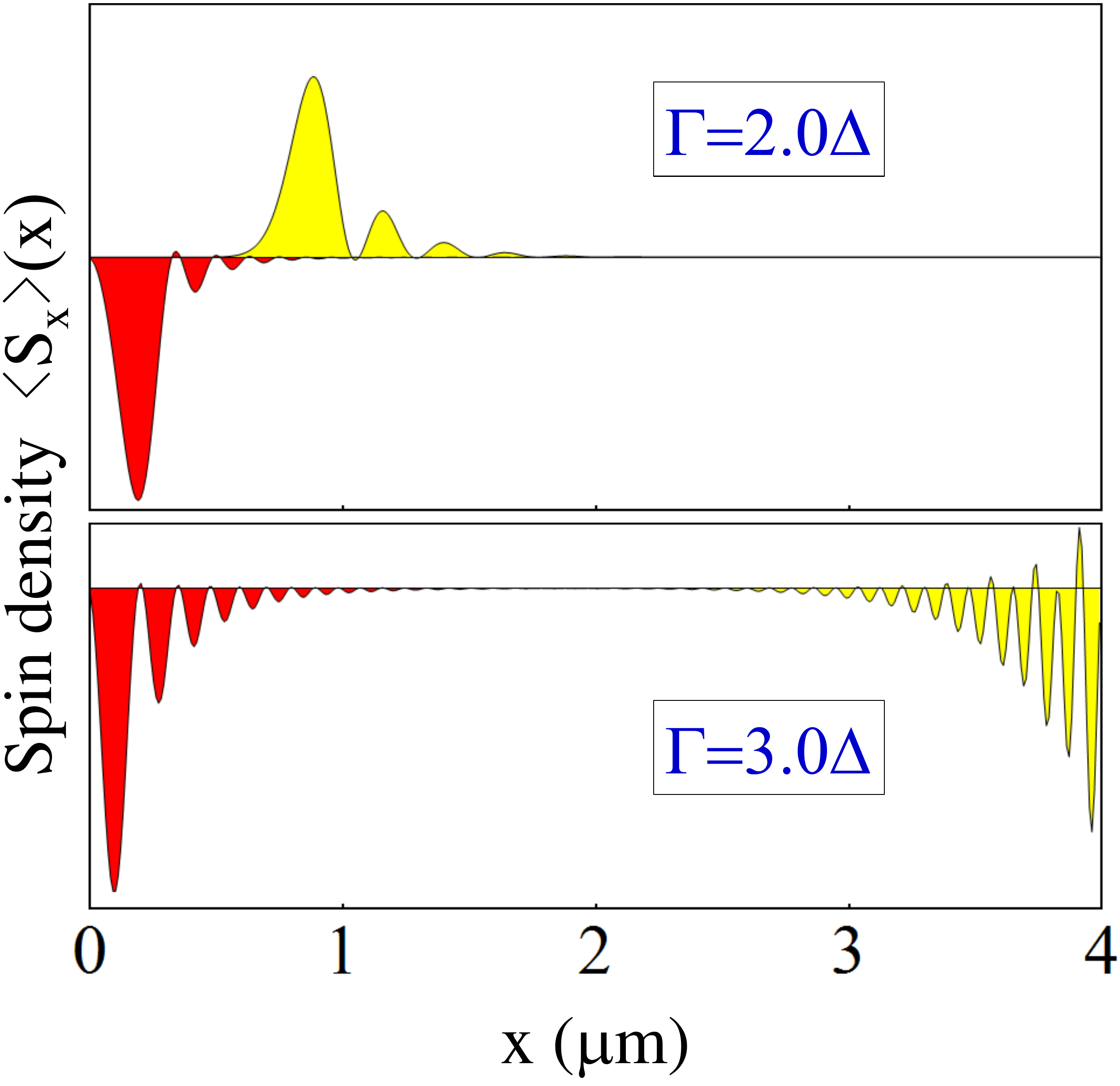}
\end{center}
\caption{Spin density of zero-energy states as defined by Eq. (\ref{Si}). In the topologically-trivial regime (upper panel) the two Majorana bound states have $x$-components of the spin density with opposite signs, revealing the fact that two Majoranas are associated with different spin-split sub-bands. By contrast, the Majorana bound states that emerge in the topological regime (lower panel) are associated with a single sub-band and have similar spin polarizations.}
\vspace{-4mm}
\label{Fig7}
\end{figure}

\subsection{Spin polarization and spin-resolved STM measurements}

In the low-tunneling limit, the spin-resolved differential conductance for tunneling into a specific small region of the wire is proportional to the spin-resolved local density of states (LDOS).  If we focus on the zero-energy spin-resolved LDOS, this is a quantity proportional to spin density $\langle{\bm S}\rangle(x)$ defined by Eq. (\ref{Si}) \cite{Simon, Kawakami, Schaffer, Schaffer_2}. Hence, a spin-resolved STM measurement of the smoothly-confined wire with the energy spectrum shown in Fig. \ref{Fig1} and the low-energy states represented in Figs. \ref{Fig2} and \ref{Fig3} will generate zero-bias spatial profiles similar to the spin density shown in Fig. \ref{Fig7}.  Note that in the topologically-trivial regime ($\gamma= 2\Delta$), the MBSs  have opposite spin polarizations along the $x$ direction (`spin' defined with only particle components of the wave function). This is a consequence of the fact that the two MBSs are associated with different spin-polarized sub bands of the semiconductor wire. A simple intuitive picture in terms of band bending due to the smooth confinement corresponds to the two spin sub-bands crossing the chemical potential at different locations along the wire, which roughly correspond to the locations of the MBSs. By contrast, in the topological regime the two MBSs have the same spin polarization, revealing the fact that they correspond to a single spin sub-band crossing the chemical potential at the opposite ends of the wire.

We conclude that spin-resolved STM can distinguish between effective MBSs emerging in the topological and the trivial superconducting regimes. However, the key to this in not so much the information associated with the spin degree of freedom, but the information associated with the spatial dependence of the measured quantity. A standard STM measurement could also distinguish between a double-peak bound state localized near the left end of the wire  (topologically trivial regime) and two bound states localized near the opposite ends of the system (topologically non-trivial regime). Moreover, tunneling into the bulk of the wire would also provide a clear signature associated with the closing of the bulk gap at $\Gamma_c$. Combined with the observation that the overlap $\Omega$ -- another quantity that contains information about the spatial profiles of the MBSs -- shows a clear signature associated with the TQPT, we conclude that some information about the spatial profiles  of the Majorana bound states is absolutely necessary to unambiguously identify them as quasiparticles emerging in a topological superconducting phase.

\subsection{Spin-resolved differential tunneling conductance} \label{SR_tunnelingB}

The most common experimental method used so far in the search for Majorana bound states is charge tunneling into the end of a superconducting wire. Unfortunately, this probe does not provide any information about the spatial profile of the MBS and, consequently, cannot clearly distinguish between the trivial and topological regimes. The simplest way to incorporate some spatial information is to perform tunneling measurements at both ends of the wire and observe correlated splitting oscillations \cite{Dassarma}. The question that we address here is whether or not performing a spin-resolved measurement  can provide additional information.

To answer this question, we take the superconducting wire with soft confinement described above and couple it at the left end to a normal lead, which is also modeled as a set of $N_y$ coupled chains. A potential barrier with a Gaussian profile is included between the wire and the normal lead and the
differential conduactance is calculated numerically for different values of the Zeeman field.

First, we consider standard charge tunneling. The results are shown in Fig. \ref{Fig8}.  A clear zero-bias peak can be observed for $\Gamma \geq 2\Delta$, but, as expected,  there is no characteristic signature associated with the TQPT at $\Gamma_c$. In other words, in the case of wire with soft confinement there is a serious danger of misinterpreting the meaning of the zero-bias peak. If in a two-terminal experiment no correlated splitting oscillations are observed, the zero-bias peak is likely associated with an effective MBS that is part of a trivial nearly-zero energy state. We note that strongly dispersing peaks in Fig. \ref{Fig8} marked by green arrows are associated with a bound state localized on the Gaussian tunnel barrier. The energy of this Andreev bound state can be modified by tuning the barrier height and its profile.

\begin{figure}[tbp]
\begin{center}
\includegraphics[width=0.48\textwidth]{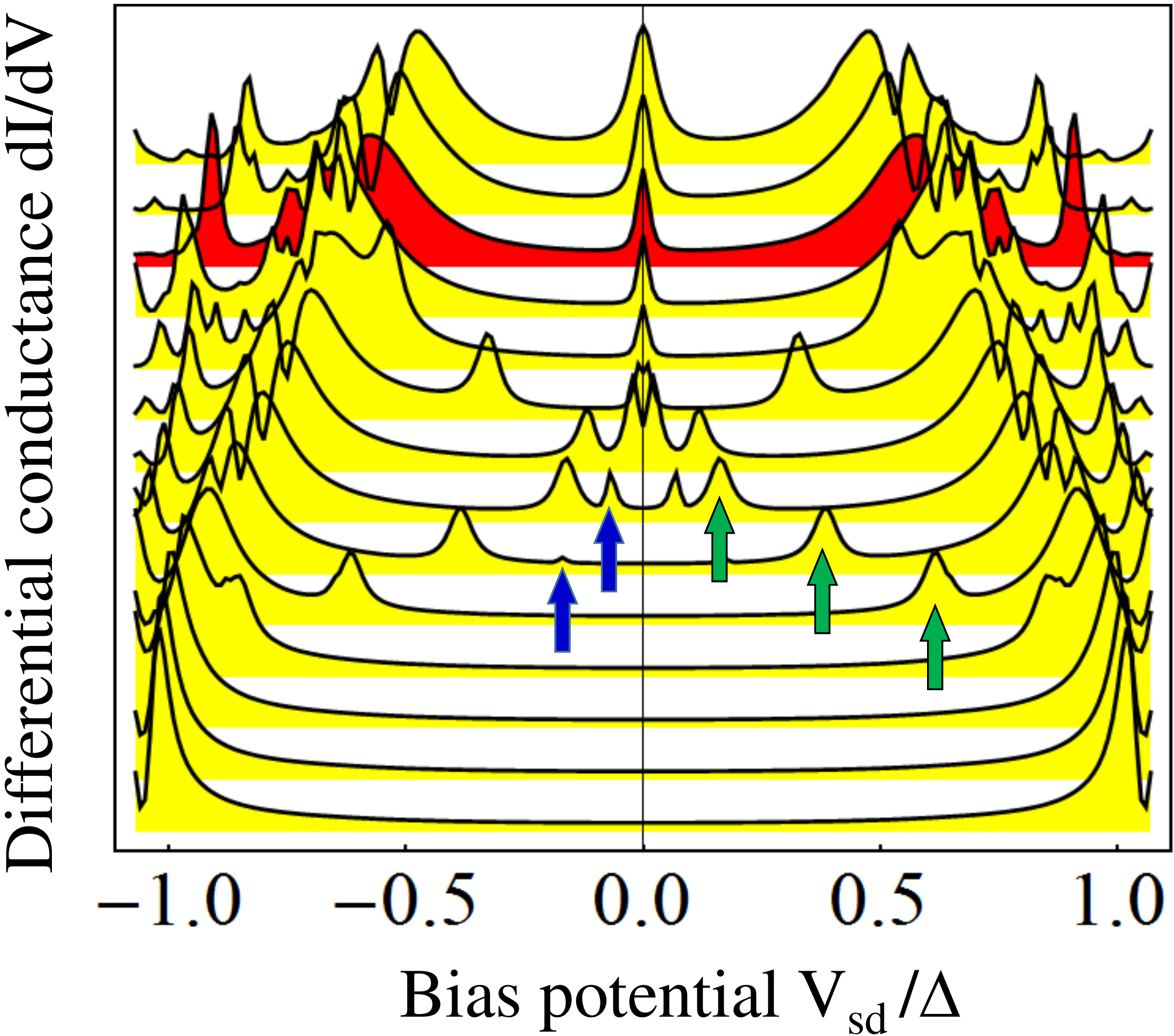}
\end{center}
\caption{Differential conductance as a function of the bias potential for different values of the Zeeman field ranging from $\Gamma=0$ (lowest curve) to $\Gamma=3.25\Delta$ (top curve) in steps of $0.25\Delta$. The curves have been shifted for clarity. The red-filled curve corresponds to $\Gamma=2.75\Delta \approx \Delta_c$; note the absence of any signature associated with the closing of the bulk gap and the emergence of a zero-bias peak for $\Gamma < \Delta_c$ (i.e. in the topologically-trivial regime). The peaks marked by blue arrows, which merge into the zero-bias peak,  correspond to the lowest energy mode (red line) in Fig. \ref{Fig1}, while the peaks marked by green arrows are associated with an Andreev bound states localized inside the potential barrier region.}
\vspace{-4mm}
\label{Fig8}
\end{figure}

Second, we consider spin-polarized tunneling. Specifically, we inject electrons with spin oriented parallel or anti-parallel to the Zeeman field $\Gamma$, i.e. along the $x$ direction. The results are shown in Fig. \ref{Fig9}.  Again, there is no signature associated with the TQPT, i.e. no possibility of distinguishing between the topological and trivial regimes. However, there is a sharp difference between two spin orientations. A spin orientation consistent with the spin polarization of the state $\psi_{01}$ shown in Fig. \ref{Fig7} generates a clearly visible zero-bias peak. By contrast, tunneling electrons with the opposite spin generates no zero-bias peak. Hence, the spin density defined by Eq, (\ref{Si}) and the corresponding spin polarization are the relevant quantities in the context of spin-resolved tunneling measurements.

As a final remark, we note that the finite-energy states with dominant contribution to $dI/dV$ are those states that have significant amplitudes at the left end of the wire. As evident from Figs. \ref{Fig8} and \ref{Fig9}, these states are weakly dependent on the applied Zeeman field, consistent with our observation related to Fig. \ref{Fig4bisbis} about the `horizontal modes' that couple to the Majorana bound state  $\psi_{01}$.

\begin{figure}[tbp]
\begin{center}
\includegraphics[width=0.48\textwidth]{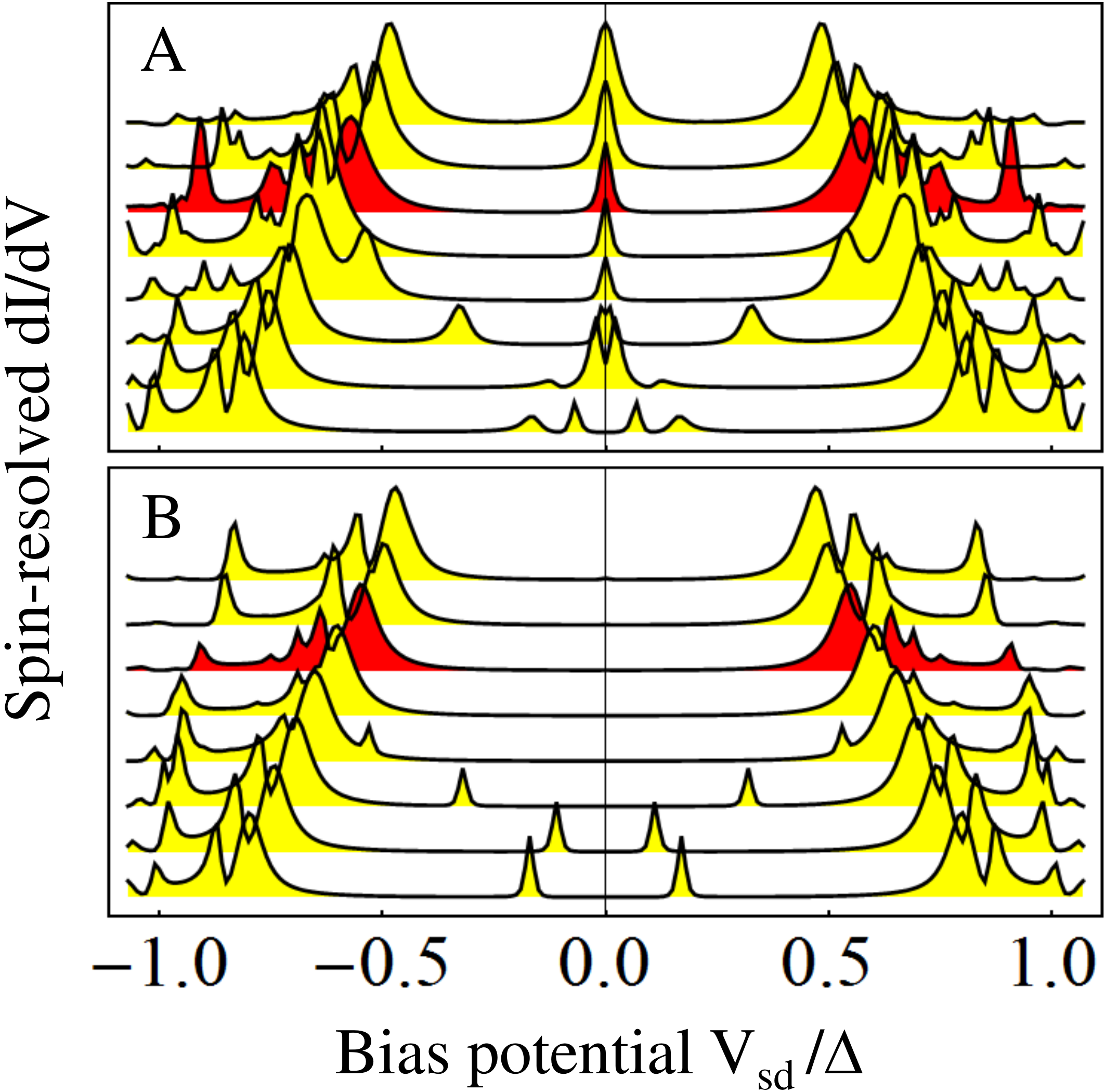}
\end{center}
\caption{Differential conductance for spin-resolved tunneling into the end of the wire. The Zeeman field ranges from $\Gamma=1.5\Delta$ (lowest curves) to $\Gamma=3.25\Delta$ (top curves) in steps of $0.25\Delta$. Panel A corresponds to the tunneling of electrons with spin parallel to the applied Zeeman field, while panel B shows the differential conductance corresponding to the opposite spin orientation. Note that that the zero bias peak is absent when tunneling electrons with a spin-orientation opposite to the spin polarization of the leftmost Majorana bound state $\psi_{01}$ (see Fig. \ref{Fig7}).}
\vspace{-4mm}
\label{Fig9}
\end{figure}

\section{Conclusions}
We discuss the concept of spin polarization of a Majorana bound state in condensed matter systems by clearly distinguishing between the average of the spin operator calculated with respect to the full Majorana fermion wave function (${\bm s}_0(i)$ defined in Eq.~(8)) and the same average calculated with respect to only the particle component of the wave function ($\langle{\bm S}\rangle(i)$ given in Eq.~(9)). We show that ${\bm s}_0(i)$, like its charge density counterpart, is identically zero for a MBS.  We emphasize that these are local (rather than global) properties: not only the total spin and charge of the Majorana bound state vanish, but the corresponding densities are also identically zero everywhere. Since it is this quantity that couples to an externally applied magnetic field, we show that neither the MBS energy eigenvalue nor the MBS wave function are affected by the external field (in a realistic system such as a semiconductor-superconductor heterostructure the change in the Majorana wave function by an applied magnetic field occurs via hybridization with the higher energy excited states, and thus disappears for a truly `stand-alone' MBS with large enough gap to the excited states). 

By an abuse of language, however, if `spin' of a MBS is defined by Eq. (9) (i.e., only with respect to the particle (or hole) components of the wave function), we show that this quantity does have a non-zero spatial profile along certain directions in spin-space. Interestingly, it is this quantity that appears in spin-resolved local density of states \cite{Simon, Kawakami, Schaffer, Schaffer_2} and thus it can be probed in spin-resolved tunneling experiments. The question of spin of a MBS, therefore, is similar to the question of its charge: although absence of coupling to local electrostatic potential supports the well-known fact that Majorana bound states carry no charge, and hence are characterized by an identically zero charge density, a Majorana bound state can in fact be probed by tunneling charge into the topological superconducting wire, the characteristic signature being a zero-bias conductance peak. By detailed numerical calculations on the semiconductor-superconductor heterostructure platform we show that no extra information can be gleaned by spin-resolved tunneling in topological superconductors (beyond what is gained by spin-unresolved tunneling) that can help us discriminate between MBSs in the topological regime and accidental zero energy states in the topologically trivial regime in the parameter space.

As an interesting byproduct of this work we also show that, in spatially inhomogeneous systems, MBSs can appear with increasing Zeeman field even in the absence of a topological quantum phase transition. When the system is still topologically trivial, a regular low energy subgap state near a soft boundary can nucleate two spatially separated zero energy states which (for all practical purposes) behave as Majorana zero modes. So long as the system remains topologically trivial, these Majorana bound states are localized inside the smooth confinement region, while in the topologically non-trivial superconducting phase (i.e., with Zeeman field larger than the critical field required for TQPT) the two  Majorana states are localized near the ends of the wire. These issues are particularly important for non-homogeneous topological superconductors, such as systems with a soft confinement as discussed here or various schemes for creating topological superfluid phases in cold atom systems. In the light of these considerations, demonstrating the nonlocal character of the topologically-protected Majorana pair and its emergence \textit{after} the system undergoes a TQPT, become critical tasks for the ongoing experimental search for Majorana bound states in solid state structures. In particular, we conjecture that observing a zero-bias conductance peak (of height $\sim 2e^2/h$) that sticks to zero energy for a certain (possibly large) range of Zeeman fields \textit{does not} represent a unique signature of the topologically protected Majorana bound states, because such a signature can also appear in the topologically trivial phase in spatially inhomogeneous systems (see Fig.~8). Ensuring the homogeneity of the chemical potential throughout the wire on an energy scale lower than the induced gap, which may be a challenging task in a setup involving multiple back-gates that control different segments of the wire, represents a critical condition for the successful realization and unambiguous demonstration of a topological superconducting phase that supports a single pair of Majorana bound states localized near the ends of the wire.

This work is supported by the National Science Foundation (grant No. DMR-1414683) and the WV Higher Education Policy Commission (Research Challenge Grant HEPC.dsr.12.29) and partially by AFOSR.


\end{document}